\documentclass[nohyper,12pt,letterpaper]{JHEP3}
\usepackage{epsfig}

\def\zzz{{\hbox{z\kern-1mm z}}}

\newcommand{\be}{\begin{equation}}
\newcommand{\ee}{\end{equation}}
\newcommand{\ben}{\begin{eqnarray}}
\newcommand{\een}{\end{eqnarray}}
\newcommand{\ba}{\begin{eqnarray}}
\newcommand{\ea}{\end{eqnarray}}
\newcommand{\nn}{\nonumber \\}

\newcommand{\beq}{\begin{equation}}
\newcommand{\eeq}{\end{equation}}

\newcommand{\bi}{\begin{itemize}}
\newcommand{\ei}{\end{itemize}}

\newcommand{\wc}{\check}

\newcommand{\vtau} {\vec \tau}
\newcommand{\vxi} {\vec \xi}

\def\ZZZ{{\hbox{ Z\kern-1.6mm Z}}}
\def\RRR{{\hbox{ R\kern-2.4mm R}}}
\def\CCC{{\hbox{ C\kern-2.0mm C}}}
\def\zzz{{\hbox{z\kern-1mm z}}}

\newcommand{\AAA}{{\cal A}}

\newcommand{\OO}{{\cal O}}

\newcommand{\LL}{{\cal L}}

\newcommand{\wt}{\widetilde}

\newcommand{\NN}{{\cal N}}

\newcommand{\lb}{\left(}
\newcommand{\rb}{\right)}

\newcommand{\sectiono}[1]{\section{#1}\setcounter{equation}{0}}

\newcommand{\p}{\partial}
\newcommand{\refb}[1]{(\ref{#1})}

\newcommand{\ltb}{\left [}
\newcommand{\rtb}{\right ]}
\newcommand{\bc}{\begin{center}}
\newcommand{\ec}{\end{center}}

\title{Asymptotic Expansion of the $\NN=4$ Dyon Degeneracy}

\author {Nabamita Banerjee, Dileep P. Jatkar, Ashoke Sen}

\abstract{We study  various aspects of
power suppressed as well as exponentially suppressed corrections 
in the asymptotic expansion of the degeneracy of quarter BPS
dyons in $\NN=4$ supersymmetric string theories. In particular
we explicitly calculate the power suppressed corrections up to second
order and  the first exponentially suppressed corrections.
We also propose a  macroscopic origin of the exponentially
suppressed corrections using the quantum
entropy function formalism. This suggests a universal pattern of
exponentially suppressed corrections to all 
extremal black hole entropies in string theory.}


\preprint{}

\begin{document}

\sectiono{Introduction and Summary} \label{intro} 

One of the major successes of string 
theory has been the matching of the
Bekenstein-Hawking entropy of a class of extremal black holes and
the statistical entropy of a system of branes carrying the same quantum
numbers as the black hole\cite{9601029}. 
The initial comparison between the
two was done in the limit of large charges. In this limit the analysis
simplifies on both sides. On the gravity side we can
restrict our analysis to two derivative terms in the action, while on
the statistical side the analysis simplifies because we can use certain
asymptotic formula to estimate the degeneracy of states for large 
charges. However given the successful matching between the
statistical entropy and Bekenstein-Hawking entropy in the large
charge limit, it is natural to explore whether the agreement continues
to hold beyond this approximation. On the gravity side this requires
taking into account the effect of higher derivative corrections and
quantum corrections in computing the entropy.
The effect of higher derivative terms is captured by the Wald's 
generalization of the Bekenstein-Hawking 
formula\cite{9307038}. 
For extremal black holes this leads to the entropy function formalism
for computing the entropy\cite{0506177}.
Recently it has
been suggested that the effect of quantum corrections to the entropy
of extremal black holes is encoded in the quantum entropy function, 
defined as the partition function of string theory on the near horizon
geometry of the black holes\cite{0809.3304}. 
On the other hand computing higher
derivative corrections to the statistical entropy requires us to compute
microscopic degeneracies of the black hole to greater accuracy. Here
significant progress has been made in a class of $\NN=4$ supersymmetric
field theories, for which we now have exact formul\ae\ for the
microscopic degeneracies\cite{9607026,0412287,0505094,
0506249,0508174,0510147,0602254,
0603066,0605210,0607155,0609109,0612011,0702141,
0702150,0705.1433,0705.3874,0706.2363,
0708.1270,0802.0544,0802.1556,0803.2692,
0806.2337,0807.4451,0808.1746,0809.4258}. (For a similar proposal in 
$\NN=2$ supersymmetric
theories, see \cite{0711.1971}.)

Our eventual goal is to compare the statistical entropy computed from the
exact degeneracy formula to the predicted result on the black hole side
from the computation of the quantum entropy function (or whatever
formula gives the exact result for the entropy of extremal black holes).
However in practice we can compute the black hole side of the result
only as an expansion in inverse powers of charges, by matching these
to an expansion in powers of derivatives / string coupling constant.
Thus we must carry out a similar expansion of the statistical entropy
if we want to compare the results on the two sides. 
A systematic procedure
for developing such an expansion of the statistical entropy has
been discussed in \cite{9607026,0412287,0510147,0605210}. 
Our main goal in this paper is to explore
this expansion in more detail, and. to whatever extent possible,
relate it to the results of macroscopic computation.

The rest of the paper is organized as follows. In \S\ref{sover}
we give a brief overview of the exact dyon degeneracy formula
in a class of $\NN=4$ supersymmetric string theories, and 
discuss the systematic procedure of  extracting the degeneracy
for large but finite charges. 
We also organise the computation of the statistical entropy by
representing the result as a sum of contributions from
single centered and multi-centered black holes, and then express the
single centered black hole entropy as an asymptotic expansion in inverse
powers of charges, together with exponentially suppressed 
corrections. In \S\ref{stat} we examine the leading exponential term
in the expression for the statistical entropy and compute the statistical
entropy to order $1/{\rm charge}^2$. Previous computation of the 
statistical entropy was carried out to order ${\rm charge}^0$.
We compare these results with the
exact result for the statistical entropy and find good agreement.
We also find that the agreement is worse if we compare the result
with the exact statistical entropy in a domain where besides single
centered black holes, we also have contribution from
two centered black holes. This confirms that the asymptotic expansion is
best suited for computing the entropy of single centered black holes.
{}From the gravity perspective these corrections
should be captured by six derivative corrections to the effective
action; however explicit analysis of such contributions has not been
carried out so far. 

In \S\ref{sub} we analyze the contribution from the
exponentially subleading terms to the entropy of single centered black
holes. While power suppressed corrections to the statistical
entropy have been compared to the higher derivative corrections
to the black hole entropy in various approximations, so far there
has been no explanation of these exponentially suppressed terms
from the black hole side.\footnote{Note that this expansion is quite
different from the Rademacher expansion studied in 
\cite{0005003,0712.0573} since we scale all the charges uniformly.}
In \S\ref{qent} we suggest a macroscopic origin of the
exponentially suppressed 
contributions to the entropy from quantum entropy function
formalism. 
In this formalism the leading contribution to the macroscopic 
degeneracy comes from path integral
over the near horizon $AdS_2$ geometry of the black hole with
appropriate boundary condition. We show that for the same boundary
conditions there are other saddle points which have different 
values of the euclidean
action. These values have precisely the form needed to reproduce
the exponentially suppressed contributions to the leading
microscopic degeneracy.

\sectiono{An Overview of Statistical Entropy Function} \label{sover}

In this section, we briefly review the systematic procedure
for computing
the asymptotic expansion of the 
statistical entropy of a dyon
in a class of $\NN=4$ supersymmetric string theories.
The approach mainly 
follows \cite{9607026,0412287,0510147,0605210,0708.1270}. 
Our notation will be that of \cite{0802.0544}.

\subsection{Dyon degeneracy}

Let us consider an $\NN=4$ supersymmetric string theory with
a rank $r$ gauge group. We shall work at a generic point in the
moduli space where the unbroken gauge group is $U(1)^r$. 
The low energy supergravity describing this theory has a
continuous
$SO(6,r-6)\times SL(2,\RRR)$ symmetry which is broken
to a discrete subgroup in the full string theory. 
We
denote by $Q$ and $P$ the $r$ dimensional electric and magnetic
charges of the theory,
by
$L$ the $SO(6, r-6)$ invariant metric and by $(Q^2,P^2,Q\cdot P)$
the combinations $(Q^TLQ,P^TLP,Q^TLP)$. Then for a fixed set of
values of discrete T-duality invariants the degeneracy $d(Q,P)$,
-- or more precisely the sixth 
helicity trace $B_6$\cite{9708062} -- of a dyon
carrying charges $(Q,P)$ is given by a formula of the form:
\begin{equation}\label{ehexp}
d(Q, P) =  (-1)^{ Q\cdot P+1}\,  {1\over a_1 a_2 a_3}
 \int _{\cal C} d\wc \rho \, d\wc\sigma \,
d\wc v \, e^{- \pi i ( \wc \rho P^2 
+ \wc \sigma Q^2 +2 \wc v Q\cdot P)}\, {1
\over \wc \Phi(\wc \rho,\wc \sigma, \wc v)}\, ,
\end{equation}
where $\wc\rho\equiv \wc\rho_1+i\wc\rho_2$, 
$\wc\sigma\equiv\wc\sigma_1+i\wc\sigma_2$ and 
$\wc v\equiv \wc v_1+i\wc v_2$ are three
complex variables, $\wc\Phi$ is a function of 
$(\wc \rho,\wc \sigma, \wc v)$ which we shall refer to as the
inverse of the dyon
partition function, and
${\cal C}$ is a three real dimensional subspace of the three
complex dimensional space labeled by $(\wc \rho,\wc \sigma,
\wc v)$, given by
\begin{eqnarray}\label{ep2int}
\wc \rho_2=M_1, \quad \wc\sigma_2 = M_2, \quad
\wc v_2 = M_3, \nonumber \\
 0\le \wc\rho_1\le a_1, \quad
0\le \wc\sigma_1\le a_2, \quad 0\le \wc v_1\le a_3\, .
\end{eqnarray}
The periods $a_1$, $a_2$ and $a_3$ of $\wc\rho$, $\wc\sigma$
and $\wc v$ are determined by the the quantization laws of
$Q^2$, $P^2$ and $Q\cdot P$.
$M_1$, $M_2$ and $M_3$ are large but fixed 
numbers.
The choice of the $M_i$'s 
depend on the domain of the asymptotic moduli space in which
we want to compute $d(Q,P)$. As we move from one domain to
another crossing the walls of marginal stability, $d(Q,P)$ changes.
However this change is captured completely by a deformation of
the contour labelled by $(M_1,M_2,M_3)$ without any change
in the partition function $\wc\Phi$\cite{0702141,0702150}. A simple
rule that expresses $(M_1,M_2,M_3)$ in terms of the
asymptotic moduli is\cite{0706.2363}:
\ben \label{echoiceint}
M_1 &=& \Lambda \, \left({|\lambda|^2\over \lambda_2} +
{Q_R^2 \over \sqrt{Q_R^2 P_R^2 - (Q_R\cdot P_R)^2}}\right)\, ,
\nonumber \\
M_2 &=& \Lambda \, \left({1\over \lambda_2} +
{P_R^2 \over \sqrt{Q_R^2 P_R^2 - (Q_R\cdot P_R)^2}}\right)\, ,
\nonumber \\
M_3 &=& -\Lambda \, \left({\lambda_1\over \lambda_2} +
{Q_R\cdot P_R \over \sqrt{Q_R^2 P_R^2 - (Q_R\cdot P_R)^2}}
\right)\, ,
\een
where $\Lambda$ is a large positive number,
\be \label{edefcrint}
Q_R^2 = Q^T (M+L)Q, \quad P_R^2=P^T (M+L)P, \quad
Q_R\cdot P_R = Q^T(M+L)P\, ,
\ee
$\lambda\equiv \lambda_1+i\lambda_2$ denotes the asymptotic value
of the axion-dilaton moduli which
belong to the gravity multiplet and $M$ is the 
asymptotic value of the $r\times r$ symmetric matrix
valued moduli field of the matter multiplet 
satisfying $MLM^T=L$.

A special point in the moduli space is the attractor point corresponding
to the charges $(Q,P)$. If we choose the asymptotic values of the 
moduli fields to be at this special point then 
all multi-centered black hole solutions are absent and
the corresponding
degeneracy formula captures the degeneracies of single centered black
hole only\cite{0706.2363}. 
This attractor point corresponds to the choice
of $(M,\lambda)$ for which
\be \label{ech1}
Q_R^2=2Q^2, \quad P_R^2=2P^2, \quad Q_R\cdot P_R=2Q\cdot P,
\quad \lambda_2 = 
{\sqrt{Q^2 P^2 - (Q\cdot P)^2}\over P^2}, \quad
\lambda_1 = {Q\cdot P\over P^2}\, .
\ee
Substituting this into \refb{echoiceint} we get
\be \label{ech2}
M_1=2\, \Lambda\, {Q^2 \over 
\sqrt{Q^2 P^2 - (Q\cdot P)^2}}, \quad
M_2 = 2\, \Lambda\, {P^2 \over 
\sqrt{Q^2 P^2 - (Q\cdot P)^2}}, \quad
M_3 = - 2\, \Lambda\, {Q\cdot P \over 
\sqrt{Q^2 P^2 - (Q\cdot P)^2}}\, .
\ee

We can invert the Fourier integrals \refb{ehexp} by writing
\begin{eqnarray}\label{phidqp} 
d(Q, P) &=& (-1)^{Q\cdot P+1}\,
g\left({1\over 2} P^2 , {1\over 2}\, Q^2, Q\cdot P\right)\, , 
\end{eqnarray}
where $g(m,n,p)$ are the coefficients of Fourier expansion of the
function $1/ \wc\Phi(\wc \rho,\wc \sigma, \wc v)$:
\begin{equation}\label{efo2} 
{1 \over \wc \Phi(\wc \rho,\wc \sigma, \wc
  v)} =\sum_{m,n,p} g(m,n,p) \, e^{2\pi i (m\, \wc \rho + n\,
  \wc\sigma + p\, \wc v)}\, .  
\end{equation}
Different choices of $(M_1,M_2,M_3)$ in \refb{ehexp} will
correspond to different ways of expanding $1/\wc\Phi$
and will lead to different $g(m,n,p)$. Conversely, for
$d(Q,P)$ associated with a given domain of the asymptotic moduli
space, if we define $g(m.n,p)$ via eq.\refb{phidqp}, then the
choice of $(M_1,M_2,M_3)$ is determined by requiring that
the series \refb{efo2} is convergent for
$(\wc\rho_2,\wc\sigma_2,\wc v_2)=(M_1,M_2,M_3)$.

A special case on which we shall focus much of our attention is
the $\NN=4$ supersymmetric string theory obtained by
compactifying type IIB string theory on $K3\times T^2$ or
equivalently heterotic string theory compactified on $T^6$.
In this case the function $\wc\Phi$ is given by the well known
Igusa cusp form of weight 10:
\begin{equation}\label{deff}
\wc\Phi(\wc \rho,\wc \sigma,\wc v )= 
\Phi_{10}(\wc \rho,\wc \sigma,\wc v )=
e^{2\pi i (\wc \rho 
+\wc\sigma
+ \wc v)}\prod_{k',l,j\in \zzz 
\atop k',l\ge 0; j<0 \, {\rm for}
\, k'=l=0}
\left( 1 - e^{2\pi i (\wc \sigma k'   + \wc \rho l + \wc v j)}
\right)^{c(4lk' - j^2)}\, ,
\end{equation}
where $c(u)$ is defined via the equation
\be \label{edefcu}
8\, \left[ {\vartheta_2(\tau,z)^2
\over \vartheta_2(\tau,0)^2} +
{\vartheta_3(\tau,z)^2\over \vartheta_3(\tau,0)^2}
+ {\vartheta_4(\tau,z)^2\over \vartheta_4(\tau,0)^2}\right]
= \sum_{j,n\in\zzz} c(4n - j^2) \, e^{2\pi i n\tau + 2\pi i j z}\, .
\ee

\subsection{Asymptotic expansion and statistical entropy
function}

In order to compare the statistical entropy 
$S_{stat}(Q,P)\equiv \ln d(Q,P)$ with
the black hole entropy we need to extract the behaviour of
$S_{stat}(Q,P)$ for large charges. We shall now briefly review
the
strategy and the results.
For details the reader is referred to \cite{0708.1270}.

\begin{enumerate}
\item Beginning with the expression for $d(Q,P)$ given in
\refb{ehexp}, we first deform the contour to small values
of $(\wc\rho_2,\wc\sigma_2,\wc v_2)$ (say of the order of
1/charge). In this case the contribution to $S_{stat}$ from the 
deformed contour can be shown to be subleading, and hence the
major contribution comes from the residue at the poles picked up
by the contour during the deformation.
\item For any given pole, one of the three integrals in
\refb{ehexp} can be done using residue theorem. The integration
over the other two variables are carried out using the method
of steepest descent. It turns out that in all known examples,
the dominant contribution to $S_{stat}$ computed using this
procedure comes from the pole of the integrand \i.e.\ zero
of $\wc\Phi$ at
\be \label{ezeropos}
\wc\rho\wc\sigma - \wc v^2 +\wc v=0\, .
\ee
Furthermore near this pole $\wc\Phi$ behaves as
\be \label{ephibehav}
\wc\Phi(\wc\rho,\wc\sigma,\wc v)
\propto (2v-\rho-\sigma)^{k} \, v^2 \, g(\rho) \, g(\sigma)\, ,
\ee
where 
\begin{equation}\label{e5nrep}
\rho 
   = {\wc \rho \wc\sigma - \wc v^2\over \wc\sigma}, 
   \qquad \sigma = {\wc\rho \wc \sigma - (\wc v - 1)^2\over  
   \wc\sigma}, \qquad
   v 
=   {\wc\rho \wc\sigma - \wc v^2 + \wc v\over \wc\sigma}\, ,
\end{equation}
$k$ is related to the rank $r$ of the gauge group via the
relation
\be \label{erank}
r = 2k + 8\, ,
\ee
and $g(\tau)$ is a known function which depends on the details
of the theory. Typically it transforms as a modular function of weight
$(k+2)$ under a certain subgroup of the $SL(2,\ZZZ)$ group.
In the $(\rho,\sigma,v)$ variables the pole at \refb{ezeropos}
is at $v=0$. The constant of proportionality in \refb{ephibehav}
depends on the specific $\NN=4$ string theory we are considering,
but can be calculated in any given theory.

\item Using the residue theorem the contribution to the integral
\refb{ehexp} from the pole at \refb{ezeropos} can be
brought to the form
\begin{equation}\label{ek1}
e^{S_{stat}(Q, P)} \equiv d(Q, P)\simeq
\int{d^2\tau\over \tau_2^2} \, e^{-F(\vec \tau)}\, ,
\end{equation}
where $\tau_1$ and $\tau_2$ are two complex variables, related
to $\rho$ and $\sigma$ via
\be \label{ereln}
\rho\equiv  \tau_1+ i \tau_2 \ \ ,\ \ \sigma\equiv -\tau_1+ i \tau_2\, ,
\ee
and
\begin{eqnarray}\label{ek2}
F(\vec\tau) &=& -\Bigg[ {\pi\over 2 \tau_2} \, |Q -\tau P|^2
-\ln g(\tau) -\ln g(-\bar\tau) - (k+2) \ln (2\tau_2)
\nonumber \\
&& +\ln\bigg\{K_0 \, \left(
2(k+3) + {\pi\over \tau_2} |Q -\tau P|^2\right)
\bigg\}\Bigg] \, , \nonumber \\
K_0 &=& constant\, .
\end{eqnarray}
Even though $\tau_1$ and $\tau_2$ are complex, we have used
the notation $\tau=\tau_1+i\tau_2$, $\bar\tau=\tau_1-i\tau_2$,
$|\tau|^2= \tau\bar\tau$,
and $|Q-\tau P|^2=(Q-\tau  P)(Q-\bar\tau P)$. 
Note that $F(\vec\tau)$ also depends on the charge 
vectors $(Q,P)$, but we have not explicitly displayed these in its
argument. The $\simeq$ in \refb{ek1} denotes equality up to
the (exponentially subleading) 
contributions from the other poles.

\item 
We can analyze the contribution to \refb{e5nrep} using the saddle
point method. To leading order the saddle point corresponds to the
extremum of the first term in the right hand side of \refb{ek2}.
This gives
\be \label{esa1}
\tau_1={Q\cdot P\over P^2}, \qquad \tau_2 = {\sqrt{Q^2P^2
-(Q\cdot P)^2}\over P^2}\, .
\ee
Using \refb{e5nrep}, \refb{ereln} we get
\be \label{esa2}
(\wc\rho, \wc \sigma, -\wc v) =  {i\over 2\sqrt{Q^2P^2
-(Q\cdot P)^2}} (Q^2, P^2, Q\cdot P)
- (0,0, {1\over 2})\, .
\ee
We can regard the result for $-S_{stat}$ as the extremal
value of the 1PI effective action in the zero dimensional
quantum field
theory, with fields $\tau,\bar\tau$ (or equivalently $\tau_1$,
$\tau_2$) and action $F(\vtau) - 2\ln \tau_2$. A manifestly
duality invariant procedure for evaluating $S_{stat}$ was
given in \cite{0605210} using background field method and Riemann
normal coordinates. The final result of this analysis is that
$S_{stat}$ is given by
\be \label{est1}
S_{stat}\simeq -\Gamma_B(\vtau_B) \qquad \hbox{at}
\qquad {\p\Gamma_B(\vtau_B)\over \p\vtau_B}=0\, ,
\ee
where $\Gamma_B(\vtau_B)$ is the sum of 
1PI vacuum diagrams calculated with the action
\be \label{est2}
\sum_{n=0}^\infty {1\over n!} (\tau_{B2})^n \xi_{i_1}\ldots
\xi_{i_n}\, D_{i_1} \cdots D_{i_n} F(\vtau)\bigg|_{\vtau=\vtau_B}
- \ln {\cal J}(\vxi) \, ,
\ee
where
\be \label{ejb}
{\cal J}(\vxi)  = \left[{1\over |\xi|}\sinh|\xi|\right]\, ,
\qquad |\xi|\equiv \sqrt{\bar \xi \xi}\, .
\ee
Here $\vtau_B$ is a fixed background value, $\xi$, $\bar\xi$ are
zero dimensional quantum fields and
\begin{eqnarray} \label{est2.5}
D_\tau (D_\tau^m D_{\bar\tau}^n F(\vec\tau))
&=& (\partial_\tau - im/\tau_2) (D_\tau^m D_{\bar\tau}^n F(\vec\tau)),
\nonumber \\
D_{\bar\tau} (D_\tau^m D_{\bar\tau}^n F(\vec\tau))
&=& (\partial_{\bar\tau} + in/\tau_2)
(D_\tau^m D_{\bar\tau}^n F(\vec\tau))\, ,
\end{eqnarray}
for any arbitrary ordering of $D_\tau$ and $D_{\bar\tau}$ in $D_\tau^m
D_{\bar\tau}^n F(\vec\tau)$.  
\end{enumerate}
This finishes the required background for generating the asymptotic
expansion of the statistical entropy to any given order in
inverse powers of charges, -- all we need is to compute 
$\Gamma_B(\tau_B)$ to the desired order and then find its value
at the extremum.
The function
$-\Gamma_B(\tau_B)$ is called the statistical entropy function.

\subsection{Exponentially suppressed corrections} \label{ssexp}

In our analysis we shall also be interested in studying the exponentially
subleading contribution to the statistical entropy.  These come from
picking up the residues at the other zeroes of $\wc\Phi$. The details
of the analysis has been reviewed in 
\cite{0708.1270}; here we summarize
the results for the special case of heterotic string 
theory on $T^6$\cite{9607026}.
In this case $k=10$, $\wc\Phi$ is given by the Siegel modular
form $\Phi_{10}$, and the periods $(a_1,a_2,a_3)$ are all equal to
1. $\Phi_{10}$ has second order zeroes at
\ben\label{ep4}
&&  n_2 ( \wc\sigma \wc\rho  -\wc v ^2) + j\wc v  
+ n_1 \wc\sigma  -m_1 \wc\rho + m_2
 = 0, \nonumber \\
&& \hbox{for} \quad  \hbox{$m_1,n_1,m_2, n_2\in\ZZZ$,
$j\in 2\ZZZ+1$}, \quad
m_1 n_1 + m_2 n_2 +\frac{j^2}{4} = {1\over 4}\, .
\een
Since eqs.\refb{ep4} are invariant under $(\vec m, \vec n, j)\to
(-\vec m, -\vec n, -j)$, we can use this symmetry to set $n_2\ge 0$.
For any given $n_2\ge 1 $ we can use 
the symmetry of $\Phi_{10}$ under
integer shifts in $(\wc\rho,\wc\sigma,\wc v)$ to bring $m_1$, $n_1$
and $j$ in the range
\be \label{epole2}
0\le n_1\le n_2-1, \qquad 0\le m_1\le n_2-1, \qquad 
0\le j\le 2n_2-1\, .
\ee
Using this symmetry we can fix $(m_1,n_1,j)$ in this range, but 
then we must extend the integration range over 
$(\rho_1,\sigma_1,v_1)$ to be over the whole real axes. 
For given $n_2$, $m_1$, $n_1$, $j$, 
the last equation in \refb{ep4} then determines $m_2$ in terms
of the other variables. This equation also forces $j$ to be odd, and 
$m_1n_1 + (j^2-1)/4$ to be an integer multiple of $n_2$.
We can now evaluate the contribution from each of these poles
using saddle point method. To leading order the location of the
saddle point from the pole associated with a given set of values of
$m_i$, $n_i$ and $j$ is given by\cite{9607026,0708.1270}
\be \label{epole4}
(\wc\rho, \wc \sigma, -\wc v) =  {i\over 2n_2 \sqrt{Q^2P^2
-(Q\cdot P)^2}} (Q^2, P^2, Q\cdot P)
-{1\over n_2}\, (n_1, -m_1, {j\over 2})\, .
\ee
For $n_2=1$ we can choose $n_1=m_1=0$, $j=1$ and \refb{epole4}
reduces to \refb{ezeropos}.

Besides these there are also contributions from the poles
corresponding to $n_2=0$. These are in fact the poles responsible
for the jump in the degeneracy as we cross walls of marginal 
stability\cite{0702141}.
In particular for the wall associated with a decay of the form
\be \label{e2t}
(Q,P) \to (Q_1,P_1)+(Q_2,P_2)\, ,
\ee
\be \label{e2.55t}
(Q_1,P_1) = (\alpha Q +\beta P, \gamma Q +\delta P), 
\qquad (Q_2,P_2)=
(\delta Q -\beta P, -\gamma Q + \alpha P)\, ,
\ee
\be \label{e3t}
\alpha\delta=\beta\gamma, \qquad \alpha+\delta=1\, ,
\ee
the jump in the index is given by the residue at the pole at
\be \label{e4t}
\wc\rho \gamma - \wc\sigma \beta + \wc v (\alpha-\delta) = 0\, .
\ee
Unlike the residues from the poles at \refb{ep4}, which grow as
exponentials of quadratic powers of charges, the residues at
the poles at \refb{e4t} grow as exponentials of linear powers of
charges. Thus one expects them to be suppressed compared to  the
contribution from all other poles of the form given in \refb{ep4}.
Nevertheless we shall see that for small charges the residues at
\refb{e4t} give substantial subleading contribution to the
statistical entropy.

\subsection{Organising the Asymptotic Expansion} \label{easym}

Consider the contour integral given in \refb{ehexp} with 
$(M_1,M_2,M_3)$ given as in \refb{echoiceint}. In order to
find the asymptotic expansion of this expression we need to deform
the contour so that it passes through the saddle point. Since the integral
is done over the real parts of $(\wc\rho,\wc\sigma,\wc v)$ keeping
their imaginary parts fixed, we shall deform the contour 
by varying the imaginary parts 
$(\wc\rho_2,\wc\sigma_2,\wc v_2)$ of $(\wc\rho,\wc\sigma,\wc v)$.
For this we first note that in the $(\wc\rho_2,\wc\sigma_2,\wc v_2)$
space, the point $(M_1,M_2, M_3)$ given in
\refb{ech2} corresponding
to the choice of the contour for single centered black holes, and the
values of $(\wc\rho_2,\wc\sigma_2,\wc v_2)$ given in
\refb{epole4} corresponding to various saddle points, lie along a
straight line passing through the origin:
\be \label{estra1}
{\wc\rho_2\over Q^2} = {\wc\sigma_2\over P^2}= -{
\wc v_2\over Q\cdot P}
\, .
\ee
Thus we can first deform the
contour from its initial position to the position \refb{ech2}, keeping
$Im(\wc\rho,\wc\sigma,\wc v)$ large all through, and then deform
it along a straight line towards the origin. In the first step we shall only
cross the poles of the type given in \refb{e4t}. This picks up the
contribution to the entropy from the multi-centered black holes
which were present at the  point in the moduli space where we are
computing the entropy. In the second stage we pick up the
contribution from all the saddle points with $n_2\ge 1$, but do not
cross any pole of the type given in \refb{e4t}. These can
then be regarded as the contribution to the entropy of a pure single
centered black hole. 
Thus we see that the complete contribution to single centered black hole
entropy comes from residues at the poles \refb{ep4} with $n_2\ge 1$.
This suggests that at least for finite values of charges
where the jumps across the walls of marginal stability are not
extremely small compared to the total index, the asymptotic expansion,
based on the residues at the poles at \refb{ep4} with $n_2\ge 1$, is
better suited for reproducing the entropy of single centered black holes
than that of single and multi-centered black holes together. We shall see
this explicitly in our numerical analysis.

\sectiono {Power Suppressed 
Corrections}\label{stat}

In \S\ref{sover} we outlined a general procedure for computing the
statistical entropy as an expansion in inverse powers of charges.
In this section we shall use this method to compute the statistical
entropy to order $1/q^2$ where $q$ stands for a generic charge. 
For comparison we note that
the leading correction to the entropy is quadratic in the charges.
Contribution to $S_{stat}$ up to order $q^0$ has been computed in 
\cite{0412287,0510147,0605210}.

We begin with the expression for $F(\vec\tau)$ given in \refb{ek2}
and carry out the background field expansion as described in
\refb{est2}.  For this we organise  
\refb{est2} as a sum of three terms
\be \label{finalf} 
F(\vtau) -\ln {\cal J}(\vxi) = F_0+F_1+F_2 
\ee 
where
\ben \label{edeffi}
F_0 &=& - {\pi\over 2 \tau_2} \, |Q -\tau P|^2\, , \nonumber \\
F_1 &=& \ln g(\tau) +\ln g(-\bar\tau) + (k+2) \ln (2\tau_2)
-\ln {\cal J}(\vxi) 
- \ln \bigg[ K_0 {\pi\over \tau_2} |Q -\tau P|^2\bigg]\, , \nonumber \\
F_2 &=& - \ln \left[1 + {2(k+3)\tau_2\over \pi |Q -\tau P|^2}\right]\, ,
\een
represent respectively 
the leading piece of order $q^2$,
the $O(q^0)$ piece and all terms
of the order $q^{-2 n}, n \geq 1$. 
Since the loop expansion is an expansion in powers of $q^{-2}$,
in order to carry out a systematic
expansion in powers of $q^{-2}$ we need to regard $F_0$ as the
tree level contribution, $F_1$ as the 1-loop contribution and $F_2$ as
two and higher loop contributions. 
To compute $\Gamma_B$ up to a
certain order, we need to compute 1PI vacuum diagrams in the zero
dimensional field theory with action $(F_0+F_1+F_2)$ up to that order
regarding $\xi$ as fundamental field. 
Thus for example in order to compute the contribution
to $\Gamma_B$ to order $q^{-2}$
we need to include all one and two loop diagrams involving vertices
from $F_0$, 
all one loop diagrams involving a single vertex of
$F_1$ and the tree level contribution from $F_0$, $F_1$ and $F_2$.

\begin{figure}
\centering
\includegraphics[height=4cm]{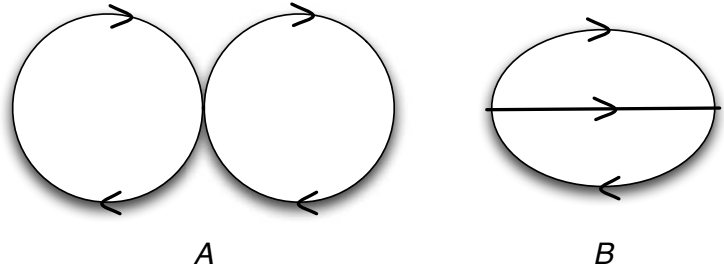}
\caption{2-loop graphs using the vertices
from $F_0$.}
\label{2loop}
\end{figure}

\begin{figure}
\centering
\includegraphics[height=4cm]{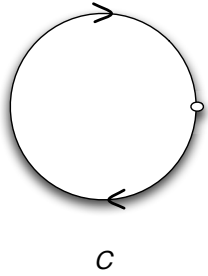}
\caption{1-loop  graph using  a 2-vertex from $F_1$.}
\label{1loop}
\end{figure}

To see more explicitly how the powers of $q$ appear,
we expand $F(\vtau)$ in field variable $\xi$ around the
background point $\vtau_B$.  We then identify the quadratic term
in $\xi$ in the leading action $F_0$ with the inverse propagator and
all other terms (including quadratic terms in the expansion of
$F_1$ and $F_2$) as vertices. 
Since $F_0$ is of order $q^2$, this gives a propagator of 
order $q^{-2}$. All 
vertices coming from $F_0$ are of order $q^2$, all vertices
coming from $F_1$ are of order $q^0$ and the vertices coming from
$F_2$ are of order $q^{-2n}$ with $n\ge 1$.
Let us now consider  a 1PI vacuum
diagram with $V_n$ number of $n$-th order 
vertices coming from $F_0$. 
Since there are no external legs, we have $\sum_n nV_n /2$ propagators.
Thus the contribution from this diagram goes as 
\be \label{ecount}
q^{\sum_n (2-n) V_n}\, .
\ee
Similar counting works for vertices coming from $F_1$ and
$F_2$, but every vertex coming from $F_1$ will carry an extra
power of $q^{-2}$ and every vertex coming from $F_2$ will carry
two or more extra powers of $q^{-2}$. Thus
an order
$q^{-2}$ contribution to the effective action can come from
\begin{equation} \label{ess2}
(V_4=1, \quad V_n=0 \quad \hbox{for}\quad n\ne 4) \quad 
\hbox{or}\quad (V_3=2, \quad 
V_n=0 \quad \hbox{for}\quad n\ne 3) \, ,
\end{equation}
if all the vertices are from $F_0$, and 
\be \label{ess1}
V_2=1, \quad V_n=0 \quad \hbox{for}\quad n\ne 2\, ,
\ee
if this single two point vertex is from $F_1$.\footnote{Note
that $F_0$ does not give a two point vertex.}
The possible diagrams associated with \refb{ess2} have been
shown in Fig.\ref{2loop} whereas the diagram associated
with \refb{ess1} have been shown in Fig.\ref{1loop}.
Finally
the order $q^{-2}$ contribution from $F_2$ is obtained by just
adding the $F_2(\tau_B)$ term to $\Gamma_B(\tau_B)$.

The above analysis shows that in order to calculate the contribution
to $\Gamma_B$ up to order $q^{-2}$, we need to expand
$F_0(\vtau)$ to quartic order in $\vec\xi$, and 
$F_1(\vtau)$ to quadratic
order in $\vec\xi$. This is done with the help of
\refb{est2}, \refb{est2.5}.
We get\footnote{Whenever a
$\tau$ ($\tau_B$) appears without a vector sign, it should be
interpreted as $\tau_1+i\tau_2$ ($\tau_{B1} + i\tau_{B2}$).}
\ben \label{egive}
F_0(\vtau) &=& F_0(\vtau_B) 
-{i\pi\over 4\tau_{B2}} \left\{ \xi (Q-\bar\tau_BP)^2 
- \bar\xi (Q-\tau_BP)^2\right\}
-{\pi\over 4 \tau_{B2}} \, |Q -\tau_B P|^2 \,
\bar\xi \xi \nonumber \\
&& +{i\pi\over 24 \tau_{B2}} \left\{ (Q-\tau_B P)^2 \bar\xi^2\xi
- (Q-\bar \tau_B P)^2 \xi^2\bar \xi\right\}
-{\pi\over 48\tau_{B2}} |Q-\tau_B P|^2 \, \bar\xi^2\xi^2\, ,
\nonumber \\
F_1(\vtau) &=& F_1(\vtau_B) +\tau_{B2} 
\left[\left\{ {g'(\tau_B)\over g(\tau_B)}
+ {k+2\over \tau_B-\bar\tau_B} +{1\over \tau_B-\bar\tau_B}\,
{(Q-\bar\tau_B P)^2 \over |Q-\tau_B P|^2}\right\} \xi + c.c.\right]
\nonumber \\
&& - \left\{ {k+4\over 4} - {(Q-\tau_B P)^2 (Q-\bar\tau_B P)^2
\over 4\, (|Q-\tau_B P|^2)^2} + {1\over 6}\right\}\, \xi\bar\xi
+\OO(\xi^2,\bar\xi^2)\, .
\een
The quadratic term in the expansion of $F_0(\vtau)$ gives the
propagator
\be \label{epropa}
M^{\xi\bar\xi}=M^{\bar\xi \xi} = -{4 \tau_{B2}\over \pi
|Q -\tau_B P|^2}, \qquad
M^{\xi\xi}=M^{\bar\xi\bar\xi} = 0\, .
\ee
Using the vertices we can evaluate the order
$q^{-2}$ contribution to $\Gamma_B$ shown in the three diagrams
in
Figs.~\ref{2loop} and \ref{1loop}. The results are
\ben \label{eabc}
A &=& -{2 \tau_{B2} \over 3 \pi |Q-\tau_B P|^2},
\nonumber \\
B 
&=&{2 \tau_{B2} (Q-\tau_B P)^2 (Q-\bar\tau_B P)^2
\over 9 \pi (|Q-\tau_BP|^2)^3}\, ,
\nonumber \\
C 
&=&{2 \tau_{B2} \over 3 \pi |Q-\tau_BP|^2}
+{(4+k)\tau_{B2} \over \pi 
 |Q-\tau_BP|^2}-{\tau_{B2} (Q-\tau_B P)^2(Q-\bar\tau_B P)^2 \over 
\pi \, (|Q-\tau_BP|^2)^3}\, .
\een
Combining this with the order $q^2$ and $q^0$ contribution to
$\Gamma_B$ given in \cite{0605210}, the complete
statistical entropy function goes as,
\begin{eqnarray}\label{entropyfunc}
\Gamma_B (\vec \tau_B)   &=&
F_0 (\vec \tau_B)+F_1(\vec \tau_B)+
F_2(\vec \tau_B) -\ln(\pi\, | M^{\xi\bar \xi}| )+A+B+C
\nn
&=& \Gamma_0 (\vec \tau_B) + 
\Gamma_1 (\vec \tau_B) + \Gamma_2 (\vec \tau_B) \nn
 \Gamma_0 (\vec \tau_B) &=& -{\pi\over 2\tau_{B2}}\, |Q-\tau_BP|^2
 ,\nn
 \Gamma_1 (\vec \tau_B) &=&  \ln g(\tau_B) +\ln g(-\bar\tau_B) 
+ (k+2) \ln (2\tau_{B2}) -\ln (4 \pi K_0) \, \nn
 \Gamma_2 (\vec \tau_B) &=& -{\tau_{B2} \over \pi |Q-\tau_B P|^2} 
 \lb(k+2)+ 
{7 \over 9} {(Q-\tau_B P)^2 (Q-\bar \tau_B P)^2 \over  
(|Q-\tau_B P|^2)^2} \rb \, .
\end{eqnarray}
The last term in $ \Gamma_2 (\vec \tau_B)$ vanishes 
at the extremum of $\Gamma_0(\vtau_B)$ where
\be \label{esaddle}
\tau_{B2} = 
{\sqrt{Q^2 P^2 - (Q\cdot P)^2}\over P^2}, \quad
\tau_{B1} = {Q\cdot P\over P^2}
\ee
We can therefore get rid of this term by doing a field
redefinition. 
Using this we can write
\be \label{egammaalt}
\Gamma_2 (\vec \tau_B) = -{\tau_{B2} \over \pi |Q-\tau_B P|^2} 
(k+2) \, .
\ee
We now note that $ \Gamma_2 (\vec
\tau_B)$ is independent of the modular form $g(\tau)$. This fact has
some important implications for our result; we will come back to it at
the end of this section.

We can now extremize $\Gamma_B(\vtau_B)$ given in
\refb{entropyfunc} with respect to $\vtau_B$
to evaluate the black-hole entropy up to this order. 
For this it is enough to find the location of the extremum to
order $1/q^2$. Let $\vtau_{(0)}$ be the extremum of 
$F_0(\vtau_B)$ given
in \refb{esaddle}. By extremizing $F_0+F_1$ we can find the
extremum to order $1/q^2$. We get
\be \label{enewsaddle}
\tau =\tau_{(0)} + {2\sqrt{Q^2P^2-(Q\cdot P)^2}\over \pi (P^2)^2}\,
\overline{{\p\Gamma_1\over \p\tau}}+\OO(1/q^4)\, ,
\ee
where the derivative of $\Gamma_1$ is taken at fixed $\bar\tau$.
Substituting this in the argument of the $\Gamma_i$'s we get
\be\label{entropy} 
S_{stat} = -\Gamma_0-\Gamma_1-\Gamma_2
= S^{(0)}+S^{(1)}+S^{(2)} ,
\ee
where 
\begin{eqnarray}\label{entropy1}
S^{(0)} &=& \pi\sqrt{Q^2P^2-(Q\cdot P)^2} \nn 
S^{(1)} &=&
-\ln g(\tau_{(0)}) -\ln g(-\bar\tau_{(0)}) - (k+2) \ln (2\tau_{(0)2}) 
+\ln(4\pi K_0) \nn
S^{(2)}&=& {2+k \over 2 \pi \sqrt{Q^2 P^2 - (Q\cdot P)^2}}
 +\ltb\lb{g'(\tau_{(0)}) \over g(\tau_{(0)})}+{k+2 
 \over \tau_{(0)}-\bar\tau_{(0)}}\rb 
\lb{g'(-\bar\tau_{(0)}) \over g(-\bar\tau_{(0)})}+{k+2 \over 
\tau_{(0)}-\bar\tau_{(0)}}\rb\rtb \nn
&& \qquad \qquad \qquad \qquad \qquad \qquad \times
 \, {4 \tau_{(0)2}^3 \over \pi |Q-\tau_{(0)} P|^2}\, .
\end{eqnarray} 

For type IIB string theory compactified on $K3\times T^2$,
$k=10$, $g(\tau)=\eta(\tau)^{24}$ and $4\pi K_0=1$.
We have shown in table \ref{t1} 
the approximate statistical entropies $S^{(0)}_{stat}=
S^{(0)}$ calculated with the `tree level' statistical entropy
function, $S^{(1)}_{stat}= S^{(0)}+S^{(1)} $ calculated with the `tree
level' plus `one loop' statistical entropy function and
$S_{stat}^{(2)}= S^{(0)}+S^{(1)}+S^{(2)} $ calculated with the `tree
level' plus `one loop' plus `two loop' statistical entropy function
and compared the results with the exact statistical entropy $S_{stat}$.
The exact results for $d(Q,P)$ are
computed using a choice of contour for which only single centered
black holes contribute to the index for $Q\cdot P>0$ and both single
and 2-centered black hole solutions contribute for $Q\cdot P<0$.
We clearly see that the asymptotic expansion has better agreement
with the exact results when only single centered black holes are
present, in accordance with our general argument. 

\begin{table} 
\begin{center}\def\st{\vrule height 3ex width 0ex}
\begin{tabular}{|l|l|l|l|l|l|l|l|l|l|} \hline 
$Q^2$ & $P^2$ & $Q\cdot P$ & $d(Q,P)$ & $S_{stat}$
& $S^{(0)}_{stat}$ & $S^{(1)}_{stat}$ & $S_{stat}^{(2)}$ 
& $D_1$ & $D_2$
\st\\[1ex]\hline \hline
2 & 2 & 0 &  $50064$ & 10.82 &  6.28
&  10.62 & 11.576  & .2 & -0.756 \st\\[1ex] \hline
4 & 4 & 0 &  $32861184$ & 17.31 &  12.57
&  16.90 & 17.382 & .41 & -0.072 \st\\[1ex] \hline
6 & 6 & 0 &  $16193130552$ & 23.51 &  18.85
&  23.19  & 23.506 & .32 & .004 \st\\[1ex] \hline
8 & 8 & 0 &  $7999169992704$ & 29.71 &  25.13
&  29.47  & 29.71 & .24 &  .000 \st\\[1ex] \hline
10 & 10 & 0 &  $4074192429737760$ & 35.943 &  31.42
&  35.754  & 35.945 & .189 &  -0.002 \st\\[1ex] \hline
6 & 6 & 1 &  $11232685725$ & 23.14 &  18.59
&  22.88  & 23.15 & .26 &  -0.01 \st\\[1ex] \hline
6 & 6 & 2 &  $4173501828$ & 22.15 &  17.77
&  21.94  & 22.198 & .21 & -0.05 \st\\[1ex] \hline
6 & 6 & 3 &  $920577636$ & 20.64 &  16.32
&  20.41  & 20.766 & .23 & -0.13 \st\\[1ex] \hline
6 & 6 & -1 &  $11890608225$ & 23.19 &   18.59
&   22.88  & 23.15 & .31 & .04  \st\\[1ex] \hline
6 & 6 & -2 &  $2857656822$ & 21.77 &  17.77
&  21.94  & 22.198 & -0.17 &  -0.43 \st\\[1ex] \hline
6 & 6 & -3 &  $2894345136$ & 21.78 &  16.32
&  20.41  & 20.766 & 1.37 & 1.01 \st\\[1ex] \hline \hline 
\end{tabular} 
\end{center}
\caption{Comparison of the exact statistical entropy to the 
tree level, one loop and two loop results obtained via the
asymptotic expansion.  
In the last two columns 
$D_1$ is the difference of the exact result and the one
loop result and $D_2$ is the difference of the exact result and the
two loop result. 
We clearly see that for $Q\cdot P>0$ where only single centered
black holes contribute to $S_{stat}$,
inclusion of the two loop results reduces the
error, at least for large charges. \label{t1}
} 
\end{table}

Given the result for the statistical entropy to this order, one would like
to see if this can be reproduced from the macroscopic calculation on the
black hole side. So far 
black hole entropy
calculation has been done for the leading supergravity action and
a subset of the four derivative terms 
which include curvature squared
contribution to the effective 
action\cite{9812082,9906094,0007195,0508042}.
The results of these two
completely independent calculations match up to 
order $q^0$ and give
us enough confidence on the expected equivalence
of the statistical entropy and the black hole entropy. However there are
many open issues. Even at the level of the four derivative terms, only
a subset of the four derivative terms have been included in the analysis
of the black hole entropy. Furthermore at 
this order the full 1PI effective action of
string theory also contains non-local terms from integrating out the
massless fermions and Wald's formula cannot even be applied in
principle to take into account the effect of these terms. Recently
a generalization of the Wald's formula for extremal black holes
in the full quantum theory has
been proposed\cite{0809.3304} (see also 
\cite{0805.0095,0806.0053}).
This will be discussed in more
detail in \S\ref{qent} in the context of exponentially suppressed
terms. However as far as the power law corrections are concerned, at
present 
we do not have
a complete
calculation of the quantum entropy function 
for quarter BPS black holes
in $\NN=4$ supersymmetric theory
even at the level of
order $q^0$ terms. This prevents us from making a concrete
statement on the agreement between the two 
entropies.\footnote{It was shown in \cite{0508174} that the
leading asymptotic expansion of the entropy
to all orders in inverse powers of charges, 
associated with the pole at
\refb{ezeropos}, is consistent
with the OSV 
formula\cite{0405146} after inclusion of certain additional
measure factors. 
Refs.\cite{talk,0601108,0808.2627} independently
derived the 
same measure factor  in the semiclassical approximation
by
requiring that the entropy is invariant under duality
transformations. 
Our goal is to derive a general formula for the
entropy of an extremal black hole based on some
principle (like AdS/CFT) from which the results of
\cite{0508174,talk,0601108,0808.2627} would follow. In particular
if one can establish that the  
asymptotic expansion of the quantum entropy
function reduces to the formula given in
\cite{0508174,talk,0601108,0808.2627}, this
will automatically prove that the quantum entropy function agrees with
the statistical entropy to all orders in inverse powers of charges.}

Given that even at order $q^0$ we do not have a complete test of
the equality between the microscopic and the macroscopic 
calculations, we cannot hope to have such a test for the order $q^{-2}$
terms calculated here. However we can say a few words about the
possible contributions on the macroscopic side which
is needed to reproduce the
order $q^{-2}$ corrections to the statistical entropy. To this end we note
that the order $q^{-2}$ correction to the statistical entropy function
$\Gamma_B(\vtau_B)$
given in \refb{egammaalt} is manifestly invariant under continuous
duality transformation
\be \label{econdu}
\tau\to {a\tau +b\over c\tau +d}, \quad \pmatrix{Q\cr P}
\to \pmatrix{a & b\cr c & d} \pmatrix{Q\cr P}\, ,
\quad ad-bc=1, \quad a,b,c,d\in \RRR\, .
\ee
Now while comparing the statistical entropy function to the black
hole entropy function, the parameters $\tau$ get identified with the
near horizon axion-dilaton modulus $\lambda$ in the heterotic
description\cite{0412287,0510147,0605210}. 
This suggests that if the required correction comes from
a local correction to the 1PI action, then the corresponding term must
be invariant under a continuous S-duality transformation. 
Furthermore since we are looking for a correction of order $q^{-2}$,
we require the correction to the Lagrangian density to be a six derivative
term. This puts a
strong restriction on the type of contribution to the local
Lagrangian density that can be responsible for such corrections. We have
not been able to find a candidate Lagrangian density. The most
straightforward method for constructing duality invariant terms
using Riemann tensors constructed out of canonical Einstein metric
does not work since all such terms vanish in the $AdS_2\times S^2$
near horizon geometry and hence do not contribute to the
entropy function to this order. 
This of course does not rule out the existence of
duality invariant terms constructed out of other fields. 
The other possibility is that these contributions
cannot be encoded in a local Lagrangian density, but come from the
non-local contributions to the quantum entropy function arising from
the path integral over string fields in
the near horizon geometry. To this end we note
that since the OSV formula reproduces the complete asymptotic
expansion to all orders in $q^{-2}$, 
if we can derive the OSV formula from the quantum entropy
function we shall automatically reproduce these corrections to the
statistical entropy.

\sectiono{Exponentially Suppressed  Corrections} \label{sub}

In this section we shall analyze the exponentially
suppressed contributions from the zeroes
of $\Phi_{10}$
given in \refb{ep4}:
\be\label{ep4rep}
n_2 ( \wc\sigma \wc\rho  -\wc v ^2) + j\wc v  
+ n_1 \wc\sigma  -m_1 \wc\rho + m_2
 = 0\, , \ee
with
\be \label{ee0}
\hbox{$m_1,n_1,m_2, n_2\in\ZZZ$,
$j\in 2\ZZZ+1$}, \quad
m_1 n_1 + m_2 n_2 +\frac{j^2}{4} = {1\over 4}\, .
\ee
For this  we define
\be\label{ee1}
\wc\Omega=\pmatrix{\wc\rho & \wc v\cr \wc v & \wc\sigma}\, ,
\ee
and look for a symplectic transformation of the form:
\be \label{ee2}
\pmatrix{\rho & v\cr v & \sigma} \equiv \Omega
= (A\wc\Omega + B) (C\wc\Omega + D)^{-1}\, ,
\ee
such that 
\be \label{ee3}
v = {n_2 ( \wc\sigma \wc\rho  -\wc v ^2) + j\wc v  
+ n_1 \wc\sigma  -m_1 \wc\rho + m_2\over \det (C\wc\Omega + D)}\, .
\ee
Here $\pmatrix{A & B\cr C & D}$ is a $4\times 4$
symplectic matrix. 
In this case \refb{ep4rep} gets mapped to $v=0$.
On the other hand the modular transformation law of
$\Phi_{10}$  gives
\be \label{ee4}
\Phi_{10}(\wc\rho,\wc\sigma, \wc v) =
\{ \det(C\wc\Omega + D)\}^{-k} \, 
\Phi_{10}(\rho,\sigma, v)\, ,
\qquad k=10\, .
\ee
Thus the behaviour of $\Phi_{10}(\wc\rho,\wc\sigma, \wc v) $ 
near the zero \refb{ep4rep}
is  given by
\be \label{ee5}
\Phi_{10}(\wc\rho,\wc\sigma, \wc v) =
- \{ \det(C\wc\Omega + D)\}^{-k} \, 4\pi^2\, v^2 \, g(\rho)\, g(v)
+\OO(v^4)\, ,
\qquad  g(\rho) = \eta(\rho)^{24}\, .
\ee
We can now substitute \refb{ee5} 
into \refb{ehexp} (with $\wc\Phi$ replaced
by $\Phi_{10}$) and evaluate the integral over $\wc v$ using
residue theorem. For this we need to regard $(\rho,\sigma,v)$
appearing in \refb{ee5} as functions of $(\wc\rho,\wc\sigma,\wc v)$
via eq.\refb{ee2}, \refb{ee3}. The result is, up to a sign,
\ben \label{ee6}
&& (-1)^{Q\cdot P}\, \int\, d\wc\rho\, d\wc\sigma\, 
e^{- \pi i ( \wc \rho P^2 
+ \wc \sigma Q^2 +2 \wc v Q\cdot P)}\, \det(C\wc\Omega + D)^{k+2}\,
(2 n_2\wc v - j)^{-2} \nn
&& \qquad \qquad \qquad \times g(\rho)^{-1}\, g(\sigma)^{-1}\, 
\left(Q\cdot P + \OO(1)\right)\, ,
\een
where $\wc v$ and
$(\rho,\sigma)$ are to be regarded as functions of 
$(\wc\rho,\wc\sigma)$ via eqs.\refb{ep4rep} and \refb{ee2}.
The last factor in \refb{ee6} proportional to $Q\cdot P$ comes
from taking the derivative of the integrand other than the pole term
with respect to $\wc v$.
We can now evaluate the $(\wc\rho,\wc\sigma)$ integral using
the saddle point method. To leading order the location of the
saddle point is obtained by extremizing the term in the exponent
of \refb{ee6} subject to the constraint \refb{ep4rep}. The
result is given in eq.\refb{epole4}:
\be \label{epole4rep}
(\wc\rho, \wc \sigma, -\wc v) =  {i\over 2n_2 \sqrt{Q^2P^2
-(Q\cdot P)^2}} (Q^2, P^2, Q\cdot P)
-{1\over n_2}\, (n_1, -m_1, {j\over 2})\, .
\ee
The result of the integration over $(\wc\rho,\wc\sigma)$ can be
expressed as
\ben \label{ee7}
&& (-1)^{Q\cdot P}\, \left[\exp\left(- \pi i ( \wc \rho P^2 
+ \wc \sigma Q^2 +2 \wc v Q\cdot P)\right)\, 
\det(C\wc\Omega + D)^{k+2}\,
(2 n_2\wc v - j)^{-2} \, g(\rho)^{-1}\, g(\sigma)^{-1}\, \right.
\nn
&& \qquad \qquad \times
\left. \left(Q\cdot P + \OO(1)\right)\, \left((\det\Delta)^{-1/2}
+ \OO(1)\right)\right]_{\rm saddle}\, ,
\een
where the subscript `saddle' denotes that we need to set $(\wc\rho,\wc
\sigma,\wc v)$ to their saddle point values given in
\refb{epole4rep}, and $\Delta$ is the $2\times 2$ matrix:
\be \label{ee8}
\Delta = i\, Q\cdot P\, 
\pmatrix{{\p^2\wc v/ \p\wc\rho^2}
& {\p^2\wc v/ \p\wc\rho\p\wc\sigma}\cr
{\p^2\wc v/ \p\wc\rho\p\wc\sigma} & 
{\p^2\wc v/ \p\wc\sigma^2}}\, .
\ee
In evaluating \refb{ee8} we need to regard $\wc v$ as a function
of $(\wc\rho,\wc\sigma)$ via eq.\refb{ep4rep}.
Explicit computation gives
\be \label{ee9}
\det\Delta = (Q\cdot P)^2\, n_2^2 / (2 n_2 \wc v-j)^4\, .
\ee
Substituting this and \refb{epole4rep}
into \refb{ee7} gives
\ben \label{ee10}
&&{1\over n_2}\exp\left(\pi \sqrt{Q^2P^2-(Q\cdot P)^2}/n_2 \right)\, 
\left[\det(C\wc\Omega + D)^{k+2}\,
g(\rho)^{-1}\, g(\sigma)^{-1}\, (1+\OO(q^{-2}))
\right]_{\rm saddle}\nn
&& \qquad \times (-1)^{Q\cdot P}\, \exp\left[ i\pi (n_1 P^2 - m_1 Q^2 + j Q\cdot P)/n_2\right]\, ,
\een
where we have how fixed the overall sign by requiring that it agrees with
the result of \cite{0708.1270} for $(m_1,n_1,n_2,m_2,j)=
(0,0,1,0,1)$.

In order to evaluate the factor $\det(C\wc\Omega + D)^{k+2}\,
g(\rho)^{-1}\, g(\sigma)^{-1}$ appearing in \refb{ee10}
explicitly, we need to find explicitly the matrix $\pmatrix{A&B\cr
C&D}$ satisfying \refb{ee3}. We shall do this explicitly for
$n_2=2$. In this case there are six possible values of $(\vec m,\vec n,
j)$ consistent with \refb{epole2}, \refb{ee0}. They are
\ben \label{esix1}
(m_1,n_1,m_2,n_2,j)&=&(0,0,0,2,1), (1,0,0,2,1), (0,1,0,2,1), \nn
&& (0,0,-1,2,3), (1,0,-1,2,3), (0,1,-1,2,3)\, .
\een
In each of these cases we can find appropriate matrices
$\pmatrix{A & B\cr C & D}$ satisfying \refb{ee3}. These 
transformations take the form:
\ben \label{emat1}
&&\Omega = \left(
\begin{array}{ll}
 \frac{\wc\rho }{(1-2 \wc v)^2-4 \wc\rho  \wc \sigma } & \frac{-2 \wc v^2+\wc v+2 \wc\rho  \wc \sigma
   }{(1-2 \wc v)^2-4 \wc\rho  \wc \sigma } \\
 \frac{-2 \wc v^2+\wc v+2 \wc\rho  \wc \sigma }{(1-2 \wc v)^2-4 \wc\rho  \wc \sigma } &
   \frac{\wc \sigma }{(1-2 \wc v)^2-4 \wc\rho  \wc \sigma }
\end{array}
\right), \quad 
\Omega = \left(
\begin{array}{ll}
 \frac{\wc\rho }{4 (\wc v-1) \wc v+2 \wc\rho -4 \wc\rho  \wc \sigma +1} & \frac{-2 \wc v^2+\wc v+\wc\rho 
   (2 \wc \sigma -1)}{4 (\wc v-1) \wc v+2 \wc\rho -4 \wc\rho  \wc \sigma +1} \\
 \frac{-2 \wc v^2+\wc v+\wc\rho  (2 \wc \sigma -1)}{4 (\wc v-1) \wc v+2 \wc\rho -4 \wc\rho  \wc \sigma +1}
   & \frac{2 (\wc v-1) \wc v+\wc\rho -2 \wc\rho  \wc \sigma +\wc \sigma }{4 (\wc v-1) \wc v+2 \wc\rho -4
   \wc\rho  \wc \sigma +1}
\end{array}
\right), \nn
&& \Omega = \left(
\begin{array}{ll}
 \frac{-2 (\wc v-1) \wc v+\wc\rho +2 \wc\rho  \wc \sigma +\wc \sigma }{(1-2 \wc v)^2-2 (2 \wc\rho +1)
   \wc \sigma } & \frac{-2 \wc v^2+\wc v+2 \wc\rho  \wc \sigma +\wc \sigma }{(1-2 \wc v)^2-2 (2 \wc\rho
   +1) \wc \sigma } \\
 \frac{-2 \wc v^2+\wc v+2 \wc\rho  \wc \sigma +\wc \sigma }{(1-2 \wc v)^2-2 (2 \wc\rho +1) \wc \sigma }
   & \frac{\wc \sigma }{(1-2 \wc v)^2-2 (2 \wc\rho +1) \wc \sigma }
\end{array}
\right),\quad
\left(
\begin{array}{ll}
 \frac{\wc\rho }{(\wc v-1)^2-\wc\rho  \wc\sigma } & \frac{1-\wc
 v}{(\wc v-1)^2-\wc\rho  \wc\sigma
   }-2 \\
 \frac{1-\wc v}{(\wc v-1)^2-\wc\rho  \wc\sigma }-2 
 & \frac{\wc\sigma }{(\wc v-1)^2-\wc\rho 
   \wc\sigma }
\end{array}
\right)\, , \nn
&& \Omega=\left(
\begin{array}{ll}
 -\frac{(1-2 \wc v)^2-4 \wc \rho  \wc \sigma }{-2 \wc v+
 \wc \rho +\wc \sigma +1} & -\frac{\wc v (2
   \wc v-3)+\wc \rho -2 \wc \rho  \wc \sigma +1}{-2 
   \wc v+\wc \rho +\wc \sigma +1} \\
 -\frac{\wc v (2 \wc v-3)+\wc \rho -2 
 \wc \rho  \wc \sigma +1}{-2 \wc v+\wc \rho +\wc \sigma +1} &
   -\frac{\wc v^2-(\wc \rho +1) \wc \sigma }{-2 \wc v+
   \wc \rho +\wc \sigma +1}
\end{array}
\right), \nn &&
\Omega = \left(
\begin{array}{ll}
 -\frac{\wc v (3 \wc v-4)-\wc \rho -3 \wc \rho  
 \wc \sigma -\wc \sigma +1}{-2 (\wc v-1) \wc v+\wc \rho +2
   \wc \rho  \wc \sigma +\wc \sigma } & \frac{\wc v-\wc \rho -1}{-2 (
   \wc v-1) \wc v+\wc \rho +2\wc  \rho 
  \wc  \sigma +\wc \sigma }+1 \\
 \frac{\wc v-\wc \rho -1}{-2 (\wc v-1) \wc v+
 \wc \rho +2 \wc \rho  \wc \sigma +\wc \sigma }+1 & \frac{-2
   \wc \rho -1}{-2 (\wc v-1) \wc v+\wc \rho +2 \wc \rho  
   \wc \sigma +\wc \sigma }+2
\end{array}
\right)
\, . \nn
\een
These transformations can be used to get $\rho$ and $\sigma$
in terms of 
$(Q^2,P^2,Q\cdot P)$ using \refb{epole4rep}. Substituting these
into \refb{ee10} and summing over the allowed values of
$(m_1,n_1,j)$ given in \refb{esix1} we get the correction to
$d(Q,P)=\exp(S_{stat})$ to this order. If we denote the resulting 
correction
to $d(Q,P)$  by $\Delta d(Q,P)$, then the values of 
$\Delta d(Q,P)$
for different values of $(Q^2,P^2,Q\cdot P)$ have been shown in
table \ref{t2}.
\begin{table} 
\begin{center}\def\st{\vrule height 3ex width 0ex}
\begin{tabular}{|l|l|l|l|l|l|l|l|l|} \hline \hline
$Q^2$ & 2 & 4 & 6 & 6 & 6 & 6  
\st\\[1ex]\hline  
$P^2$ & 2 & 4 & 6 & 6 & 6 & 6 \st\\[1ex] \hline
$Q\cdot P$  & 0 &  0 & 0 & 1 & 2 & 3  \st\\[1ex] \hline
$\Delta d(Q,P)$ & 34.617  & 480.638  & 18537.1   
& 20104.8 & 27652.3 & 0  \st\\[1ex] \hline
\hline  
\end{tabular} 
\end{center}
\caption{First exponentially suppressed contribution to $d(Q,P)$
and $S_{stat}(Q,P)$. Note that the correction vanishes
accidentally for $Q\cdot P=Q^2/2=P^2/2$ odd. \label{t2}
} 
\end{table}

\sectiono{Macroscopic Origin of the
Exponentially Suppressed Corrections} \label{qent}

We have seen that the corrections to the leading contribution to the
statistical entropy are of two types, power suppressed corrections
which arise from expansion about the saddle point associated with
pole \refb{ezeropos}, and exponentially suppressed 
corrections associated
with the contribution from the residues at the other 
poles \refb{ep4}.
Given that we have not been able to reproduce even the power 
suppressed corrections from the macroscopic side, it 
may seem futile to
attempt to understand the exponentially suppressed corrections. However
we shall now argue that quantum entropy function may provide a
natural mechanism for understanding the exponentially suppressed
corrections.

We shall begin with a lightening review of the quantum entropy
function. 
Let us consider an extremal black hole with an $AdS_2$ factor in the
near horizon geometry. We shall regard string theory in this background
as a two dimensional theory, treating all other directions as
compact.
The background fields describing the
$AdS_2$ near horizon geometry has the form\cite{9812073}
\be \label{et2}
ds^2
= v\left(-(r^2-1) dt^2+{dr^2\over
r^2-1}\right),  \quad F^{(i)}_{rt} = e_i, \quad \cdots
\ee
where $F^{(i)}_{\mu\nu} = \p_\mu A^{(i)}_\nu - 
\p_\nu A^{(i)}_\mu$ are the gauge field strengths associated with
two dimensional gauge fields $A_\mu^{(i)}$, $v$ and $e_i$ are
constants and $\cdots$ denotes near horizon values of other fields.
Under euclidean continuation
\be \label{et6}
t = -i\theta\, ,
\ee
we have
\be \label{et6.5}
ds^2
= v\left((r^2-1) d\theta^2+{dr^2\over
r^2-1}\right),  \quad F^{(i)}_{r\theta} = -i \, e_i, \quad \cdots
\ee
Under a further coordinate change
\be \label{et6.6}
r = \cosh\eta\, \, ,
\ee
\refb{et6.5} 
takes the form
\be \label{et7}
{ds^2} = { v \, \left(d\eta^2 +\sinh^2\eta \, d\theta^2 \right),}
\qquad
F^{(i)}_{\theta\eta} = i e_i \, \sinh\eta, \qquad \cdots \, .
\nonumber \ee
The metric is non-singular at the
point $\eta=0$ if we choose $\theta$ to have period $2\pi$.
Integrating the field strength we can get the form of the gauge
field:
\be \label{et8}
A_\mu^{(i)} dx^\mu = -i \, e_i \, (\cosh\eta  
{ -1}) d\theta=  -i \, e_i \, (r  
{ -1})d\theta\, .
\ee
Note that the $-1$ factor inside the parenthesis is required to make the
gauge fields non-singular at $\eta=0$.  In writing \refb{et8} we have
chosen $A^{(i)}_\eta=0$ gauge. If $q_i$ denotes the charge 
of the black hole corresponding
to the $i$th gauge field and $\LL$ denotes the Lagrangian density
evaluated in the near horizon geometry \refb{et7}, then $\vec q$
and $\vec e$ are related as
\be \label{echre}
q_i={\p (v\LL)\over \p e_i}\, .
\ee

Quantum entropy function is a proposal for computing the exact
degeneracy of  states of an
extremal black hole. It is given by
\be \label{ealt}
d(\vec q) = \left\langle \exp[-
i  q_i\ointop d\theta \, A^{(i)}_\theta]
\right\rangle^{finite}_{AdS_2}\, ,
\ee
where $\langle ~\rangle_{AdS_2}$ 
denotes the unnormalized path integral
over various fields of string theory on euclidean global
$AdS_2$ described in \refb{et7} and $A^{(i)}_\theta$
denotes the component of the $i$-th gauge field along the boundary
of $AdS_2$.
The superscript `${finite}$' refers to the finite part of the
amplitude defined as follows.
If we regularize the infra-red divergence by putting an
explicit cut-off that regularizes the volume of $AdS_2$, then
the amplitude 
has the form $e^{C L}\times$ a finite part where $C$ is a
constant and $L$ is the length of the boundary of regulated
$AdS_2$. We define the finite part as the one obtained by
dropping the $e^{CL}$ part.
This equation gives a precise relation
between the microscopic degeneracy and an
appropriate partition function in
the near horizon geometry of the black hole.

In defining the path integral over $AdS_2$ we need to put boundary
conditions on various fields. 
We require that the asymptotic geometry coincides with \refb{et7}.
Special care is needed to fix the boundary
condition on $A^{(i)}_\theta$.
In the $A^{(i)}_\eta=0$ gauge
the Maxwell's equation around
this background has two independent solutions near the boundary: 
$A^{(i)}_\theta
={\rm constant}$ and $A^{(i)}_\theta \propto r$. Since the latter
is the dominant mode we put boundary condition on the latter mode,
allowing the constant mode of the gauge field to fluctuate.
This corresponds to working with fixed asymptotic values of the
electric fields, or equivalently fixed charges via eq.\refb{echre}.

Let us now review how in the classical limit the quantum entropy
function reduces to the exponential of the Wald entropy.  
For this we need to put an infra-red cut-off; this is done by
restricting the coordinate $r$ in the range $1\le r\le r_0$.
Then 
in the classical limit the quantum entropy
function is given by the finite part of
\be \label{ecc1}
\exp\left(-A_{bulk} - A_{boundary} - i q_i \, 
\ointop A^{(i)}_\theta \, d\theta\right)\, ,
\ee
where $A_{bulk}$ and $A_{boundary}$ represent contributions from
the bulk and the boundary terms in the classical
action in the background
\refb{et7}.
If $\LL$ denotes the
Lagrangian density of the two dimensional theory, 
then the bulk contribution to the action
in the background \refb{et7} takes the form:
\ben \label{ea2}
A_{bulk} &=& - \int d^2 x \,  
\sqrt{\det g} \, \LL \nonumber \\
&=&    -\int_0^{2\pi} d\theta \, \int_0^{\cosh^{-1}r_0 }\,
d \eta \, \sinh\eta\, v \, \LL \nonumber \\
&=& - 2\pi \, v \, \LL \, (r_0-1) +\OO(r_0^{-1})\, .
\een
In going from the second to the third step in \refb{ea2} we have
used the fact that due to the $SO(2,1)$ invariance of the $AdS_2$
background, $\LL$ must be independent of $\eta$ and $\theta$.
In this parametrization the length  $L$  of the boundary is
given by
\be \label{ea3}
L = 
\sqrt v\, \int_0^{2\pi} \sqrt{r_0^2  - 1} \, d\theta =
2\pi\, \sqrt v\, r_0  + \OO(r_0^{-1})\, .
\ee
The contribution from the last term in \refb{ecc1} can also be
calculated easily using the expression for $A^{(i)}_\theta$
given in \refb{et8}. We get
\be \label{ea3.5}
i q_i \, 
\ointop A^{(i)}_\theta \, d\theta
= 2\pi\, \vec q\cdot \vec e (r_0-1)\, .
\ee
Finally,
the contribution from $A_{boundary}$ 
can be shown to have the form\cite{0809.3304}
\be \label{eend1}
A_{boundary} = 2\pi r_0 \, K + \OO(r_0^{-1})\, ,
\ee
for some constant $K$. This gives
\ben \label{eend3}
\exp\left(-A_{bulk} - A_{boundary} - i q_i \, 
\ointop A^{(i)}_\theta \, d\theta\right)
&=& \exp\left [-2\pi r_0 (\vec q\cdot \vec e - v\, \LL + K)+
\OO(r_0^{-1})\right]\nonumber \\
&& \times
\exp\left[2\pi (\vec q\cdot \vec e - v\, \LL) \right]\, .
\een
Thus the quantum entropy function, given by the finite part of
\refb{eend3}, takes the form
\be \label{eend4}
d(q) \simeq \exp\left[2\pi (\vec q\cdot \vec e - v\, \LL) \right]\, .
\ee
The right hand side of \refb{eend4} is the exponential of the
Wald entropy\cite{0506177}.\footnote{For the special case of two
derivative actions this has also been noted recently in
\cite{0809.4264}.}
For the particular case of quarter
BPS black holes in $\NN=4$ supersymmetric string theories the
leading contribution to \refb{eend4}
has the form
\be \label{en=4}
d(q)\simeq \exp\left(\pi\sqrt{Q^2P^2-(Q\cdot P)^2}\right)\, .
\ee

Quantum corrections to \refb{eend4} can be of two types. First of
all we can have fluctuations of the string field around the
$AdS_2$ background \refb{et6.5}.
We expect this to produce power law corrections, but not change
the exponent in \refb{en=4} which is related to the finite
part of the action in the $AdS_2$ background. The other class of
corrections could come from picking altogether different 
classical 
solutions with the same asymptotic field configuration as the one
given in \refb{et6.5}. These could have different actions and hence
give contributions with different exponential factors. Thus such
corrections are the ideal candidates for producing exponentially
subleading corrections to the degeneracy.

Can we identify classical solutions which could produce the
subleading corrections discussed in \S\ref{sub}? To this end consider
a $\ZZZ_N$ quotient of the background \refb{et6.5} by the
transformation
\be \label{eq1}
\theta\to \theta+{2\pi\over N}\, .
\ee
If we denote by $(\wt r, \wt\theta)$ the coordinates of this
new space then the
solution may be expressed as
\be \label{eq2}
ds^2
= v\left((\wt r^2-1) d\wt\theta^2+{d\wt r^2\over
\wt r^2-1}\right),  \quad F^{(i)}_{\wt r\wt\theta} = -i \, e_i, \quad \cdots,
\qquad \wt\theta\equiv \wt\theta+{2\pi\over N}\, .
\ee
Since $\wt\theta$ has a different period than $\theta$,
this does not manifestly 
have the same asymptotic form as the solution
\refb{et6.5}. Let us now make a change of coordinates
\be \label{eq3}
 r = \wt r/N, \quad \theta =N\wt \theta\, .
\ee
In this coordinate system the new metric takes the form:
\be \label{eq4}
ds^2
= v\left((r^2-N^{-2}) d \theta^2+{dr^2\over
r^2-N^{-2}}\right),  
\quad F^{(i)}_{r \theta} = -i \, e_i, \quad \cdots,
\qquad \theta\equiv \theta+{2\pi}\, .
\ee
This has the same asymptotic behaviour as the original solution
and hence is a potential saddle point that could contribute to the
quantum entropy function. The action associated with this solution,
with the cut-off $r\le r_0$, can be easily calculated. 
After removing the $r_0$ dependent piece we get the
following classical contribution to the quantum entropy 
function\footnote{This is easiest to derive in the $(\wt r,\wt\theta)$
coordinate system where the total action is $1/N$ times the
action for the original $AdS_2$ background with $r_0$ replaced
by $\wt r_0$. Since $\wt r_0 = N r_0$, the terms linear in $r_0$
are the same as in the original $AdS_2$ background, whereas the
$r_0$ independent term gets divided by $N$.}
\be \label{eq5}
\exp\left[2\pi (\vec q\cdot \vec e - v\, \LL)/N \right]
= \exp\left(\pi\sqrt{Q^2P^2-(Q\cdot P)^2}/N\right)\, .
\ee
This has precisely the right form as the exponentially subleading
contributions described in \S\ref{sub} if we identify $N$ with the
integer $n_2$ appearing there. 

This however cannot be the complete story. From the form of the
solution given in \refb{eq2} it is clear that the the solution
has a $\ZZZ_N$ orbifold singularity of the type $\RRR^2/\ZZZ_N$
at the origin $\wt r=1$. This is {\it a priori} a singular configuration and
it is not clear if this is an allowed configuration in string theory.
We resolve this difficulty by accompanying the $\ZZZ_N$ action
by an internal $\ZZZ_N$ transformation
\be \label{eq6}
\phi\to \phi-{2\pi\over N}\, ,
\ee
where $\phi$ is the azimuthal coordinate of the sphere $S^2$ that
is also part of the near horizon geometry of the black hole. If $\psi$
denotes the polar angle on $S^2$ then 
the orbifold group has fixed points at $(\wt r=1, \psi=0)$ and
$(\wt r=1, \psi=\pi)$. Thus the manifold is still singular but now
the singularities are of the type $\CCC^2/\ZZZ_N$, and these can
certainly be resolved in string theory. Thus we conclude that the
resulting configuration is non-singular. The classical action is
not affected by the additional shifts in the $\phi$ coordinate and hence
the contribution to the quantum entropy function continues to be
given by \refb{eq5}.

There is however a new issue that we need to address. Now the
identification $\theta\equiv \theta+2\pi$ changes to
\be \label{eq7}
(\theta,\phi)\equiv \left(\theta+2\pi, \phi-{2\pi\over N}\right)\, .
\ee
Thus one needs to check if this is consistent with the asymptotic
boundary conditions imposed on various fields. To this end we note
that if we denote by $\AAA_\mu$ the two dimensional gauge field
arising from the $\phi$ translation isometry, then the twisted
boundary condition \refb{eq8} is equivalent to switching on 
a Wilson line of the form
\be \label{eq8}
\ointop \AAA_\theta \, d\theta = {2\pi\over N}\, .
\ee
Now as discussed earlier, for all gauge fields the boundary conditions
fix the electric field, or equivalently the charge, but the zero modes
of the gauge fields are allowed to fluctuate. Here the charge associated
with the gauge field $\AAA_\mu$ is the angular 
momentum\cite{0606244} which has
been taken to be zero. But there is no constraint on the Wilson line
$\ointop \AAA_\theta \, d\theta$. Thus we are instructed to integrate
over different possible values of this Wilson line, and in that process
pick up contribution from the different saddle points given in
\refb{eq4}. This shows that there is no conflict between the
asymptotic boundary conditions and the twist described in
\refb{eq7}.

Another issue that needs attention is integration over bosonic and
fermonic zero modes associated with this solution.
The near horizon geometry of the black hole has 
an $\NN=4$ superconformal
algebra. The generators of this algebra are the $SL(2,R)$ generators
$L_0$, $L_{\pm 1}$, the $SU(2)$ generators $J^3$, $J^\pm$ and
the supersymmetry generators 
$G^{\pm\alpha}_{\pm {1\over 2}}$.
with $\alpha=1,2$.  Of these $(L_1-L_{-1})/2$ is the
generator of rotation about the origin of $AdS_2$ and $J^3$ is the
generator of rotation about the north pole of $S^2$. Since  
the orbifold action is generated by
$(L_1-L_{-1}-2J^3)$, the quotient is not invariant under the full
$\NN=4$ superconformal algebra; it is invariant only under a
subalgebra that commutes with $(L_1-L_{-1}-2J^3)$. 
This subalgebra is 
generated by $L_1-L_{-1}$, $J^3$,  $G^{+\alpha}_{1/2}
+G^{+\alpha}_{-1/2}$ and $G^{-\alpha}_{1/2}
-G^{-\alpha}_{-1/2}$. The broken bosonic and fermionic generators
leads to four bosonic and four fermionic zero modes of the solution.
Of these the bosonic zero modes parametrize
the coset $(SL(2,R)/U(1))\times (SU(2)/U(1))=AdS_2\times S^2$.
This is precisely the situation analyzed in 
\cite{0608021}.\footnote{The
notation of \cite{0608021} is slightly different; what we are calling
$L_1-L_{-1}$ was called $L_0$ in \cite{0608021}.}
Naively the integration over the bosonic zero modes will produce
infinite result and the fermionic zero mode integrals vanish. But it
was shown in \cite{0608021} that we 
can regularize the inregrals by adding
to the action an extra term that does not affect the integral. 
The extra term lifts both the bosonic and the fermionic zero modes
and as a result the path integral produces a finite result.

There are  several other minor issues which need to be addressed. 
For type II string
theory in flat space-time, the $\ZZZ_N$ orbifold action described here
generates an allowed configuration. Here we have an $AdS_2\times
S^2$ background instead of flat space. Hence the original analysis
is not strictly valid.
However  since the orbifold fixed
point is localized in $AdS_2\times S^2$, it should not `feel' the
effect of the background geomery and continue to be
an allowed configuration. What is not guaranteed is that
the blow up modes which allow us to deform the configuration away
from the orbifold point will remain flat directions. This is an
important issue we need to address if we want to explore 
the constant multiplying \refb{eq5}. 
We also need to explore if there can be any additional
contribution to the
action from the orbifold fixed point. We expect however that since
the fixed point is localized at a point in $AdS_2\times S^2$, to
leading order such a contribution (if non-zero)
will be independent of the background
geometry of $AdS_2\times S^2$. In particular it will not have a
factor proportional to the size of $AdS_2\times S^2$, and hence will
at most give an order $q^0$
correction to the leading term $\pi\sqrt{Q^2P^2-(Q\cdot P)^2}/N$
in the exponent of \refb{eq5}. 

The analysis described above is independent of which kind of extremal
black hole we are considering.\footnote{For higher dimensional
black holes the near horizon geometry contains
a (squashed) $S^n$ factor instead of $S^2$. In this case we can 
choose a suitable embedding of
the $\ZZZ_N$ action inside the symmetry group of (squashed)
$S^n$.}
This suggests a universal pattern of the
exponentially suppressed corrections to the entropy of 
all 
extremal black
holes.  If we
denote by $S_0$ the leading contribution to the entropy then the 
exact degeneracy should contain subleading corrections of order
$e^{S_0/N}$ for all $N\in \ZZZ$, $N\ge 2$. It will be interesting to see
if the exact degeneracy formul\ae\ of extremal black holes in theories 
with less number of supersymmetries obey this structure.

\vspace{1cm}
\noindent
{\large {\bf Acknowledgement}}\\

We would like to thank Justin David, Suvankar Dutta, 
Shamik Banerjee, Debashis Ghoshal,
Rajesh Gopakumar, Rajesh Gupta,
Suvrat Raju and Sumathi Rao for useful discussions.

\end{document}